# A High-Order Sliding Mode Observer: Torpedo Guidance Application

*Ahmed Rhif, Zohra Kardous, Naceur BenHadj Braiek*

Advanced System Laboratory, Polytechnic School of Tunisia
(Laboratoire des Systèmes Avancés (L.S.A), Ecole Polytechnique de Tunisie, BP 743, 2078 La Marsa, Tunisie)

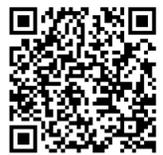

## ABSTRACT

The guidance of a torpedo represents a hard task because of the smooth nonlinear aspect of this system and because of the extreme external disturbances. The torpedo guidance reposes on the speed and the position control. In fact, the control approach which is very solicited for the electromechanical systems is the sliding mode control (SMC) which proved its effectiveness through the different studies. The SMC is robust versus disturbances and model uncertainties; however, a sharp discontinuous control is needed which induces the chattering phenomenon. The angular velocity measurement is a hard task because of the high level of disturbances. In this way, the sliding mode observer could be a solution for the velocity estimation instead of a sensor. This article deals with torpedo guidance by SMC to reach the desired path in a short time and with high precision quality. Simulation results show that this control strategy and observer can attain excellent control performances with no chattering problem.

## I. INTRODUCTION

Most modern torpedoes are completely autonomous. They have active sonar which gives them abilities to direct themselves to the target which have been designated before launch. Other types of torpedoes, for example, self-possessed, have an acoustic sensor (passive sonar) that allowed them to track the noise emitted by the engines of the target. Modern torpedoes are powered by steam or electricity, they have speeds ranging from 25 to 45 knots, and their scope ranges are from 4 to 27 km. They consist of four elements: the warhead, the air section, rear section, and tail section. The warhead is filled with explosive (181 to 363 kg). In a torpedo, steam-air section is about one third of the torpedo and contains compressed air and fuel tanks and water for the propulsion system. In fact, the torpedo guidance reposes on the speed and the position control. In this way, the sliding mode control (SMC) has largely proved its effectiveness through the reported theoretical studies; its principal scopes of application are robotics, mobile vehicle, and the electrical engines [1-4]. The advantage of such a control is its robustness and its effectiveness through the disturbances and the uncertainties of the model.

Indeed, to make certain the convergence of the system to the wished state, a high level control in addition to a discontinuous control is often requested. This fact generates the chattering phenomenon which can be harmful for the actuators. In fact, there are many solutions suggested to this problem. In literature, (SMC) with limiting band has been considered. This solution consists in replacing the discontinuous part of the control by a saturation function. Also, fuzzy control was proposed as a solution, thanks to its robustness. In another hand, the high-order sliding mode consists in the sliding variable system derivation. This method allows the rejection of the chattering phenomenon while maintaining the robustness of the approach. The high-order SMC can be represented by two algorithms:

- The twisting algorithm: the system control is increased by a nominal control $u_e$; the system error, on the phase plane, rotates around the origin until been cancelled. If we derive the sliding surface (S) n times we notice that the convergence of the state to the sliding surface *S* is even more accurate when *n* is higher.
- The super twisting algorithm: the system control is composed of two parts $u_1$ and $u_2$ with $u_1$ equivalent control and $u_2$ the discontinuous control used to reject disturbances. In this case, to obtain a sliding mode of order *n*, we have to derive the error of the system n times [5].

In the literature, different approaches have been proposed for the synthesis of nonlinear surfaces. In Ref. [6], the proposed area consists of two terms, a linear term that is defined by the Herwitz stability criteria and another nonlinear term used to improve transient performance. In Ref. [7], to measure the armature current of a DC motor, Zhang Li used the high-order sliding mode because it is faster than traditional approaches such as vector control.... To eliminate the static error that appears when measuring parameters, one uses a PI controller [8-14]. Thus, the authors have chosen to write the sliding surface in a transfer function of a proportional integral form while respecting the convergence properties of the system to this surface. The same problem of the static error was treated by adding an integrator block just after the SMC [15].

## 2. PROCESS MODELING

Torpedoes [Figure 1] are systems with strong non linearity and always subject to disturbances and model parameter uncertainties which makes their measurement and their control a hard task. Equation (1) represents the torpedo's motion's dynamic equation in 6 degrees of freedom. $M$ is the matrix of inertia and added inertia, $C$ is the matrix of Coriolis and centrifugal terms, $D$ is the matrix of hydrodynamic damping terms, $G$ is the vector of gravity and buoyant forces, and $\tau$ is the control input vector describing the efforts acting on the torpedo in the body-fixed frame. $B$ is a nonlinear function depending of the actuators characteristics, and $u$ is the control input vector [16].

$$M\dot{v} + C(v)V + S(v)v + G(\eta) = \tau$$
$$\tau = B(u) \quad (1)$$

For the modelling of this system, two references are defined [Figure 1]: one fix reference related to the torpedo which defined in an origin point: $R_0 (X_0, Y_0, Z_0)$ and the second one related to the Earth $R(x, y, z)$.
where
$\omega$ is linear velocity, $q$ the angular velocity, $\theta$ the angle of inclination, and $z$ the depth. The system control is provided by $u$ which presents the immersion deflection.

The torpedo presents a strong nonlinear aspect that appears when we describe the system in three dimensions (3D), so the state function will present a new term of disturbances $\varphi$ as shown in (2).
$$\dot{X} = AX + Bu + \varphi(X,u) \quad (2)$$
with $\| \varphi(X,u) \| < MX$ where M>0.
As we consider only the linear movement in immersion phase, we need only four degrees of freedom, for that we describe the system only in two dimensions (2D). All development done, the resulting state space describing the system is given by (3).
$$\dot{X} = AX + Bu \quad (3)$$
The system could be represented by two parts [17]: $H_1(p)$ the transfer function of inclination (4) and $H_2(p)$ the transfer function of immersion (5).

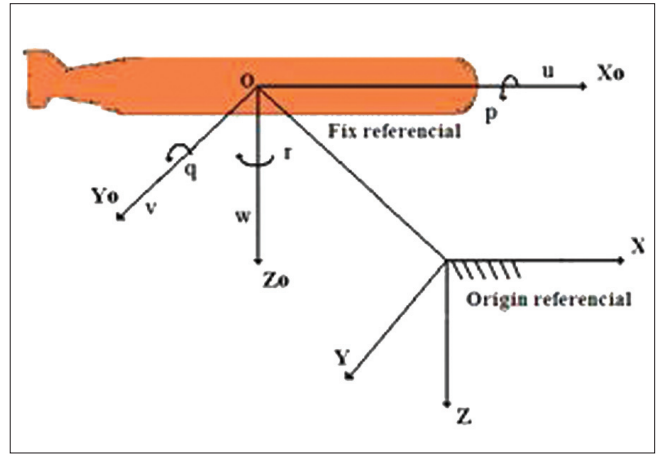

**Figure 1:** Inertial frame and body-fixed frame

$$H_1(p) = \frac{7660}{P(P)+40} \quad (4)$$

$$H_2(p) = \frac{6514(p+6.85)}{P(P+1.91)(P+12.5)(P+40)} \quad (5)$$

## 3. THE CONTROLLER DESIGN

### 3.1 The SMC
The SMC consists in bringing back the state trajectory toward the sliding surface and to make it move above this surface until reaching the equilibrium point. The sliding mode exists when commutations between two controls $u_{max}$ and $u_{min}$ remain until reaching the desired state. In another hand, the sliding mode exists when: $s\dot{s} < 0$. This condition is based on Lyapunov quadratic function. In fact, control algorithms based on Lyapunov's method have proven effectiveness for controlling linear and nonlinear systems [18].

To ensure the existence of the sliding mode, we must produce a high level commutation control. For that we will use a relay which commutates between two extreme values of control. Second, we have to define a first-order sliding surface. In this case we will choose the surface form written in (6).
$$s(t) = k_1 e(t) + k_2 \dot{e}(t) \quad (6)$$
In the convergence phase to the sliding surface, we

have to verify that:
$$\dot{V} = \frac{\partial}{\partial t^2}(s^2) \leq -\eta|s| \tag{7}$$
with η>0
In this case, the control law of the sliding mode is given by (8).
$$u = kSign(s) \tag{8}$$
with:
$$sign(s) = \begin{cases} 1, s > 0 \\ -1, s < 0 \end{cases} \tag{9}$$
*sign(.)* is the sign function and *k* a positive constant that represents the discontinuous control gain

*Chattering phenomenon*
The SMC has always been considered a very efficient approach. However, considered that it requires a high-level frequency of commutation between two different control values, it may be difficult to put it in practice.

In fact, for any control device which presents non linearity such as delay or hysteresis, limited frequency commutation is often imposed, other ways the state oscillation will be preserved even in vicinity of the sliding surface. This behavior is known by chattering phenomenon.

This highly undesirable behavior may excite the high-frequency unmodeled dynamics which could result in unforeseen instability and can cause damage to actuators or to the plant itself. In this case, the high-order sliding mode can be a solution.

### 3.2 High-order SMC synthesis
The aim of the high-order SMC is to force the system trajectories to reach in finite time the sliding ensemble of order $r \geq p$ defined by:
$$S^r = \{x \in IR^n : s = \dot{s} = ... = s^{(r-1)} = 0\}, r \in IN \tag{10}$$
*p>0*, *s(x,t)* the sliding function: it is a differentiable function with its *(r - 1)* first-time derivatives depending only on the state *x(t)* (that means they contain no discontinuities) [19-21]. In the case of second-order SMC, the following relation must be verified:
$$s(t,x) = \dot{s}(t,x) = 0 \tag{11}$$
The derivative of the sliding function is
$$\frac{d}{dt}s(t,x) = \frac{\partial}{\partial t}s(t,x) + \frac{\partial}{\partial x}s(t,x)\frac{\partial x}{\partial t} \tag{12}$$
Considering relation (12), the following equation can be written
$$\dot{s}(t,x,u) = \frac{\partial}{\partial t}s(t,x) + \frac{\partial}{\partial x}s(t,x)\dot{x}(t) \tag{13}$$
The second-order derivative of *S(t,x)* is:

$$\frac{d^2}{dt}s(t,x,u) = \frac{\partial}{\partial t}\dot{s}(t,x,u) + \frac{\partial}{\partial x}\dot{s}(t,x,u)\frac{\partial x}{\partial t} + \frac{\partial}{\partial u}\dot{s}(t,x,u)\frac{\partial u}{\partial t} \tag{14}$$

This last equation can be written as follows:
$$\frac{\partial}{\partial t}\dot{s}(t,x,u) = \xi(t,x) + \Psi(t,x)\dot{u}(t) \tag{15}$$
with:
$$\xi(t,x) = \frac{\partial}{\partial t}\dot{s}(t,x,u) + \frac{\partial}{\partial t}\dot{s}(t,x,u)\dot{x}(t) \tag{16}$$
$$\Psi(t,x) = \frac{\partial}{\partial u}\dot{s}(t,x,u) \tag{17}$$
We consider a new system whose state variables are the sliding function *s(t,x)* and its derivative $\dot{s}(t,x)$.
$$\begin{cases} y_1(t,x) = s(t,x) \\ y_2(t,x) = \dot{s}(t,x) \end{cases} \tag{18}$$

Eqs. (15) and (18) lead to (19)
$$\begin{cases} \dot{\omega}_1(t,x) = \omega_2(t,x) \\ \dot{\omega}_2(t,x) = \xi(t,x) + \Psi(t,x)\dot{u}(t) \end{cases} \tag{19}$$

In this way, a new sliding function $\sigma(t,x)$ is proposed:
$$\sigma(t,x) = \alpha_2\omega_2(t,x) + \alpha_1\omega_1(t,x) = \alpha_2\dot{s}(t,x) + \alpha_1 s(t,x) \tag{20}$$
with $\alpha_1, \alpha_2 > 0$
Eqs. (6) and (19) lead to (20)
$$\sigma = \beta_1 e + \beta_2 \dot{e} + \beta_3 \ddot{e} \tag{21}$$
with $\beta_1, \beta_2, \beta_3, > 0$

### 3.3 The sliding mode observer
The angular velocity measurements of the torpedo represent a hard task and could give unreliable results because of the very high disturbances that influence the tachymetry sensor. In this way, it may be adequate to estimate the angular velocity $\Omega$ using an observer instead of the sensor. This operation would give better results and will reduce the number of the sensors which are very costly and may have hard maintenance skills.

The observer can reconstruct the state of a system from the measurement of inputs and outputs. It is used when all or part of the state vector cannot be measured. It allows the estimation of unknown parameters or variables of a system. This observer can be used to reconstruct the speed of an electric motor, for example, from the electromagnetic torque. It also allows reconstructing the flow of the machine etc.…
The observed velocity error is noted in (22).

$$\varepsilon_\Omega = \Omega - \hat{\Omega} \qquad (22)$$

with $\hat{\Omega}$ the estimated angular velocity.

In this case, the system (3) could be represented as follow:

$$\dot{X} = AX + Bu + \lambda sign(\varepsilon_\Omega) \qquad (23)$$

with $\lambda > 0$.

## 4. SIMULATION RESULTS

The simulation results of systems (4) and (5) are given by the flowing figures [Figures 2-5].

Figures 2 and 3 show that using PID controller, the system does not reach the desired value and have a very sharp oscillations frequency. In the case of the first-order sliding mode control (SM1), we can reach the desired value in a short time, other ways we notice that the reaching phase presents some oscillations known as chattering effect. In another hand, the second-order sliding mode (SMC2) reduces considerably the chattering phenomenon and the convergence time to the desired state while preserving the robustness aspect of the control approach.

The inclination $\theta$ of the torpedo is close to zero and its evolution is more stable in the case of the second-order SMC which gives more stability effect to the system [Figure 3].

Finally, we can conclude the excellent estimation quality of the sliding mode observer [Figures 4 and 5] which illustrates that the estimated velocity and current of the control are very close to the real measurements.

## 5. CONCLUSION

In this work, torpedo controllers have been presented, detailed, and justified by simulation results. We approached the synthesis method of a control law by sliding mode using a nonlinear sliding surface. In the first time, we presented the class and the properties of this sliding surface adopted. Then, a SMC using the sliding surface developed together with stability studies were elaborated. Finally, second-order of SMC was developed with a sliding mode observer and then tested by simulation on a torpedo. Simulation results show the ability of the SMC for the tracking process of the needed path and the high estimation quality of the observer.

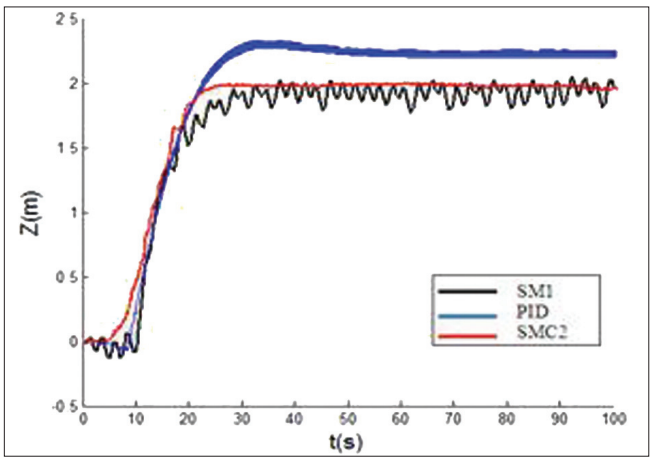

**Figure 2:** System immersion

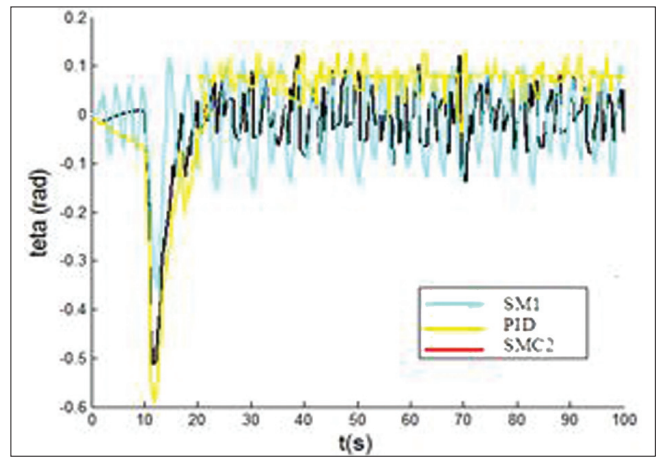

**Figure 3:** System inclination

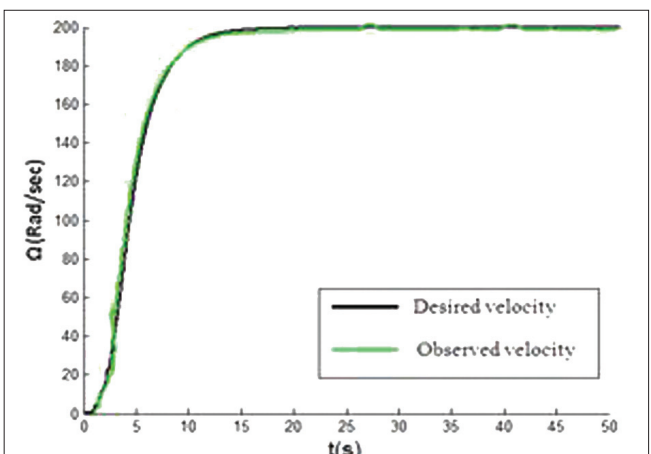

**Figure 4:** Desired and observed velocity

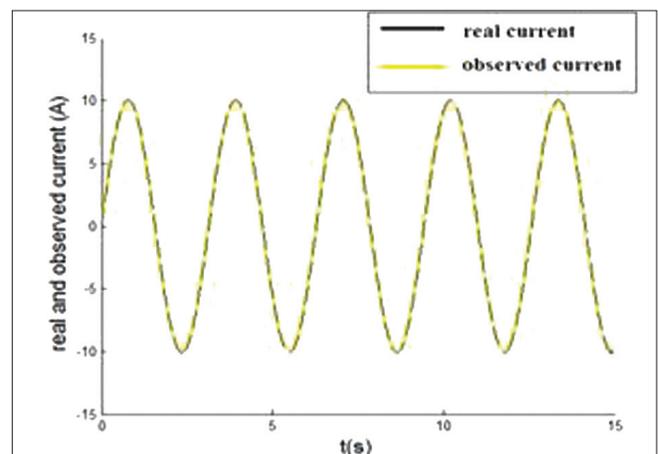

**Figure 5:** Real and observed current of the control

## REFERENCES

1. D. V. Anosov, "On stability of equilibrium points of relay systems," Automation and Remote Control, vol. 2, pp. 135-149, 1959.
2. S. V. Emel'yanov, "On pecularities of variables structure control systems with discontinuous switching functions," Doklady ANSSR, vol. 153, pp. 776-778, 1963.
3. D. Boukhetala, F. Boudjema, T. Madani, M. S. Boucherit, and N. K. M'Sirdi, "A new decentralized variable structure control for robot manipulators," Int. J. of Robotics and Automation, vol. 18, pp. 28-40, 2003.
4. D. S. Lee, and M. J. Youn, "Controller design of variable structure systems with nonlinear sliding surface," Electronics Letters, vol. 25, no. 25, pp. 1715-1716, 1989.
5. M. Rolink, T. Boukhobza, and D. Sauter, "High order sliding mode observer for fault actuator estimation and its application to the three tanks benchmark", Author manuscript, vol. 1, 2006.
6. A. Boubakir, F. Boudjema, C. Boubakir, and N. Ikhlef, « Loi de Commande par Mode de Glissement avec Une Surface de Glissement Non Linéaire Appliquée au Système Hydraulique à Réservoirs Couplés », 4th International Conference on Computer Integrated Manufacturing CIP, 03-04 November 2007.
7. Z. Li, and Q. Shui-sheng, "Analysis and Experimental Study of Proportional-Integral Sliding Mode Control for DC/DC Converter", Journal of Electronic Science and Technology of China, vol. 3, no.2, 2005.
8. B. Singh and V. Rajagopal, "Decoupled Solid State Controller for Asynchronous Generator in Pico-hydro Power Generation", IETE Journal of Research, Vol. 56, pp. 139-15, 2010.
9. B. Singh and G. Kasal, "An improved electronic load controller for an isolated asynchronous generator feeding 3-phase 4-wire loads", IETE Journal of Research, Vol. 54, pp. 244-254, 2008.
10. E. O. Hernandez-Martinez, A. Medina and D. Olguin-Salinas, "Fast Time Domain Periodic Steady-State Solution of Nonlinear Electric Networks Containing DVRs", IETE Journal of Research, Vol. 57, pp.105-110, 2011.
11. A. Rhif, "Stabilizing Sliding Mode Control Design and Application for a DC Motor: Speed Control", International Journal of Instrumentation and Control Systems, Vol. 2, pp. 25-33, 2012.
12. A. Alaybeyoglu, K. Ericiyes, A. Kantarci, and O. Dagdeviren, "Tracking fast moving targets in wieless sensor networks", IETE Technical Review, Vol. 27, pp. 46-53, 2010.
13. S. P. Hsieh, T. S. Hwang and C. W. Ni, "Twin VCM controller deseign for the nutator system with evolutionary algorithme", IETE Technical Review, Vol. 26, pp 290-302, 2009.
14. N. Petrellis, N. Konofaos, and G. Ph. Alexiou, "Estimation of a target position based on infrared pattern reception quality", IETE Technical Review, Vol. 27, pp. 36-45, 2010.
15. A. Rhif, "A High Order Sliding Mode Control with PID Sliding Surface: Simulation on a Torpedo", International Journal of Information Technology, Control and Automation, Vol.2, pp.1-13, 2012.
16. Cyrille Vuilme, "A MIMO Backstepping Control with Acceleration Feedback for Torpedo", Proceedings of the 38th Southeastern Symposium on System Theory, USA, pp. 157-162, 2006.
17. S. Que Nam, S. Hong Kown, W. Suck Yoo, M. Hyung Lee, and W. Soo Jeon, "Robust fuzzy control of o three fin torpedo", Journal of the society of naval architechts of japan, Vol.173, pp. 231-235, Korea, 1993.
18. A. Rhif, Z. Kardous, and N. Ben Hadj Braiek, "A high order sliding mode-multimodel control of non linear system simulation on a submarine mobile", Eigth International Multi-Conference on Systems, Signals & Devices, Tunisia, pp.14-19, 2011.
19. A. Rhif, "Position Control Review for a Photovoltaic System: Dual Axis Sun Tracker", IETE Technical Review, Vol. 28, pp. 479-485, 2011.
20. A. Rhif, "A Review Note for Position Control of an Autonomous Underwater Vehicle", IETE Technical Review, Vol. 28, pp. 486-493, 2011.
21. A. Rhif, "A Sliding Mode Control with a PID sliding surface for Power output Maximizing of a Wind Turbine," STM, Journal of Instrumentation and Control, Vol.2, pp. 11-16, 2012.

## Author's Biography

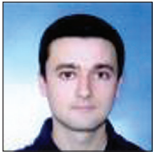

**Ahmed Rhif** was born in Sousah, Tunisia, in August 1983. He received his Engineering Diploma and Master Degree, respectively, in Electrical Engineering in 2007 and in Automatic and Signal Processing in 2009 from the National School of Engineer of Tunis, Tunisia (E.N.I.T). He has worked as a Technical Responsible and as a Project Manager in both LEONI and CABLITEC (Engineering automobile companies). Then he has worked as a researcher Assistant at the Private University of Sousah (U.P.S) and now in the High Institute of Applied Sciences and Technologies of Sousah (I.S.S.A.T.so). He is currently pursuing his PhD degree in the research laboratory "Laboratory for Advanced Systems" at the Polytechnic School of Tunisia (E.P.T). His research interest includes control and nonlinear systems.

E-mail: ahmed.rhif@gmail.com

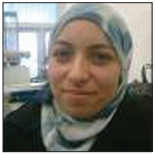

**Zohra Kardous** was born in 1976 in Tunis-Tunisia. She obtained the Engineering Diploma and Master degree, respectively, in Electrical Engineering in 1999 and in Automatic and Signal Processing in 2000 from the National School of Engineer of Tunis (ENIT), then the PhD degree in electrical engineering from both the Central School of Lille (EC-Lille) and the National School of Engineers of Tunis (ENIT) in 2004. She joined the National School of Engineers of Tunis as an Assistant Professor in 2005. Her current research interest includes modeling, control and stability analysis of nonlinear dynamical systems.

E-mail: zohra.kardous@enit.rnu.tn

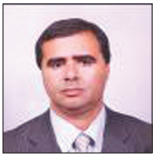

**Naceur BenHadj Braiek** was born in 1963 in Mahdia-Tunisia. He obtained the Engineering Master Degree from the National School of Engineers of Tunis (ENIT) in 1987, the PhD from the University of Science and Technology of Lille- France in 1990 and the Science State Doctorate (Doctorat d'Etat dès Sciences) from the National School of Engineers of Tunis (ENIT) in 1995, all in electrical engineering. Currently he is professor at the High School of Sciences and Techniques of Tunis (ESSTT), University of Tunis, and head of the research laboratory "Laboratory for Advanced Systems" at the Tunisia Polytechnic School (EPT). His area of interest includes modeling, analysis, control and optimization of nonlinear dynamical systems on both theoretical developments and applications.

E-mail: naceur.benhadj@ept.rnu.tn

Announcement

### "Quick Response Code" link for full text articles

The journal issue has a unique new feature for reaching to the journal's website without typing a single letter. Each article on its first page has a "Quick Response Code". Using any mobile or other hand-held device with camera and GPRS/other internet source, one can reach to the full text of that particular article on the journal's website. Start a QR-code reading software (see list of free applications from http://tinyurl.com/yzlh2tc) and point the camera to the QR-code printed in the journal. It will automatically take you to the HTML full text of that article. One can also use a desktop or laptop with web camera for similar functionality. See http://tinyurl.com/2bw7fn3 or http://tinyurl.com/3ysr3me for the free applications.